\documentclass[twocolumn,twoside,slac_two]{revtex4}
\usepackage{graphicx}
\usepackage{fancyhdr}
\newcommand{\iab}{\ensuremath{\mathrm{ab}^{-1}}}
\newcommand{\BBbar}{\ensuremath{\mathrm{B\overline{B}}}}
\newcommand{\DDbar}{\ensuremath{\mathrm{D\overline{D}}}}
\newcommand{\elec}{\ensuremath{\mathrm{e}^-}}
\newcommand{\pos}{\ensuremath{\mathrm{e}^+}}
\newcommand{\epem}{\pos\elec}
\newcommand{\prot}{\ensuremath{\mathrm{p}}}
\newcommand{\aprot}{\ensuremath{\overline{\mathrm{p}}}}
\newcommand{\ppbar}{\prot\aprot}
\newcommand{\bz}{\ensuremath{\mathrm{B}^0}}
\newcommand{\dstar}{\ensuremath{\mathrm{D}^*}}
\newcommand{\dz}{\ensuremath{\mathrm{D}^0}}
\newcommand{\dzbar}{\ensuremath{\overline{\mathrm{D}}{}^0}}
\newcommand{\dplus}{\ensuremath{\mathrm{D}^+}}
\newcommand{\ds}{\ensuremath{\mathrm{D}_s}}
\newcommand{\kp}{\ensuremath{\mathrm{K}^+}}
\newcommand{\km}{\ensuremath{\mathrm{K}^-}}
\newcommand{\qqbar}{\ensuremath{q\overline{q}}}

\newenvironment{chair}{\itshape\noindent}{}
\newcommand{\question}[1]{\bfseries\itshape\boldmath #1 \mdseries\unboldmath}

\begin{document}

\title{The future of charm physics: a discussion}

\author{B.D.~Yabsley (ed.)}
\affiliation{School of Physics, University of Sydney. NSW 2006, AUSTRALIA.}

\begin{abstract}
We closed the CHARM 2007 workshop with a lively panel discussion on the future of the field.
This document presents a summary of the key points,
and a lightly edited transcript of the discussion itself.
\end{abstract}

\maketitle

\thispagestyle{fancy}

\section{Introduction}

The CHARM 2007 organisers (and advisory committee)
deliberately chose to close the workshop 
not with a summary talk, but with a panel discussion. 
This document, accordingly,
tries to respect the idiosyncrasy of that discussion,
rather than seeking to be comprehensive or to be too ``on-message''.

The panel discussion followed four talks on future facilities
--- the relevant speakers were all panel members --- 
and could helpfully be read together with those presentations.
The raw material for the discussion was a set of questions from workshop
attendees, and some absent committee members.

\section{Key points}

\begin{itemize}
  \item	Charm physics is intellectually alive as a field,
	and we plan to keep it that way.
\end{itemize}

\textbf{Mixing and CPV}

\begin{itemize}
  \item	We now have good evidence for mixing,
	and may even breach the $5\sigma$ threshold in individual channels,
	with the $2\,\iab$ that's currently within reach. 
	With twice that dataset, the chance will be good.
	(And CDF will help!)

  \item	There's no obstacle (in principle) to calculating potential
	beyond-the-Standard-Model contributions to mixing on the lattice.
	But current methods won't cope with the nonlocal operators
	that we believe dominate the SM contribution.

  \item	Finding and studying CP violation in the charm sector is (we hope)
	the next big thing. CPV-in-mixing has long been a goal,
	but is only now within reach, with evidence for mixing established;
	controlling systematics at the $10^{-3}$ -- $10^{-4}$ level
	may be challenging.  And let's not forget direct CPV studies \ldots
\end{itemize}

\textbf{QCD, spectroscopy, lattice \ldots}

\begin{itemize}
  \item	The charm sector is a good laboratory for QCD;
	we're still learning from it, and continued to do so at this workshop.

  \item	Not every bump is a new state:
	it's possible to get complicated structure from a single pole.
	Beyond the mere ``discovery'' of ``new states'', 
	experimental determination of quantum numbers and decay properties
	is important to provide enough information for theoretical analysis.
	Interpretation can be difficult even so.

  \item	Very large data samples, and improvement in quality
	(\emph{e.g.}\ BES-II $\to$ BES-III), will help.

  \item	A really good precision test of lattice QCD requires the use
	of CKM-free quantities, \emph{e.g.}\ the ratio of semileptonic to
	leptonic decay rates. Very large data samples will help here, too.

  \item	Experimentalists and non-lattice theorists
	should interpret any lattice calculation in an informed critical spirit.
	Is the calculation unquenched? With $2+1$ sea quarks?
	How thorough is the systematic error analysis?
	If the result is to be compared to experiment,
	there should be a positive answer on all three fronts.
\end{itemize}

\textbf{Facilities in the future}

\begin{itemize}
  \item	There are plans at BaBar to maintain an analysis effort (including
	priority-setting) after data-taking stops. Still, it may be tough.

  \item	There should be active experiments in the flavour sector, 
	to complement the LHC.

  \item	There is strong support for a so-called super-B or super-flavour
	factory in our community, in addition to BES-III.
	While the B-physics programme is its selling
	point, it will be highly capable in charm
	(\& charmonium \& ISR \& \ldots)

  \item	The way forward in hadronic \ppbar\ physics is to take a global
	approach, with a full-capability detector: PANDA.
	(Earlier-generation experiments were already at their limits.)
	
  \item	\textbf{It would be very helpful to form a ``shopping list'' of
	measurements we want LHCb to make in the charm sector.}
	It could have real influence. Analysis will be limited by manpower
	and interest as much as the trigger \ldots\ and note that the upgrade
	will likely have a software trigger, to the benefit of charm studies.
\end{itemize}

\section{Dramatis personae}

\noindent
\textbf{Panel members:}
\begin{itemize}
  \item[{[DA]}] David Asner
  \item[{[DB]}]	Diego Bettoni
  \item[{[AE]}] Aida El-Khadra
  \item[{[EG]}]	Eugene Golowich
  \item[{[PS]}] Patrick Spradlin 
  \item[{[YW]}] Yifang Wang
  \item[{\itshape and}]
		{\itshape Bruce Yabsley \\
		 (chair: comments listed in italics)}
\end{itemize}

\noindent
\textbf{Contributors from the floor:}
\begin{itemize}
  \item[{[EE]}] Estia Eichten
  \item[{[BM]}] Brian Meadows
  \item[{[KP]}] Klaus Peters
  \item[{[AP]}] Alexey Petrov
  \item[{[JR]}] Jonas Rademaker
  \item[{[AS]}] Alan Schwartz
  \item[{[KS]}] Kamal Seth
  \item[{[??]}]	at least one significant contributor is currently anonymous
\end{itemize}

\section{The discussion}

\begin{chair}%
Thank you everyone for joining us for this last session.
You know everyone on the panel: they don't need any further
introduction. Let me just show you what we're going to do. You can see
the outline here, divided into
{\bf\itshape
\begin{itemize}
  \item	current issues:
	\begin{itemize}
	  \item	mixing and CP violation, and
	  \item	everything else;
	\end{itemize}
  \item	the future of the field:
	\begin{itemize}
	  \item	in general, and
	  \item	concerning future facilities.
	\end{itemize}
\end{itemize}
}
\noindent
We'll try to give about 15 minutes
on each of those categories, picking out questions out as we go.
But to begin, I'm going to throw the microphone to Gene: we've heard
four talks from experimentalists, I thought we'd get a few remarks from
the other side of the divide; and then we'll head to questions.
\end{chair}

\subsection{Overview}
\label{disc-gene}

\begin{itemize}
  \item[{[EG]}]
	Bruce asked me to be provocative and nasty, but it's really not in my
	nature.  Actually over lunch we were talking about the future of the
	field, and I was drifting off, and ended up in a fantasy world where
	things were done the right way. And in this world the LHC was in fact
	built and came on the air, and found the Higgs, and found many new
	events that we couldn't explain with the Standard Model. And people
	had realised that in order to interpret these possible signals of new
	physics, we would also have to have flavour
	physics studies of rare phenomena, so that we could start to see
	patterns emerging \ldots\ and working symbiotically together, the LHC
	and the flavor sector would get to the root of what was happening,
	something that would be very difficult if not impossible to do with
	the LHC alone.

	But then I woke up. And I thought about a colloquium I'd given
	recently, where one of the chief experimentalists there took me into
	his office and shut the door and said to my face, ``Flavor physics is
	dead!'' and apparently he's not the only one who said it: some pretty
	important people have said it. And when something like that is said
	over and over it begins to have a truth of itself.

	So, anyone who's been at this workshop realises that that's not the
	case: that not only are we getting wonderful results, there are all
	sorts of ideas --- arguments even --- about what to do for the future,
	about what's going on, and that's the sign of an intellectually alive
	effort. And part of what we're doing on this panel, I think, is trying
	to see whether or not we can realise the potential of the field.

	I know that in my own shopping-list I tend to divide it between
	electroweak physics and QCD, and I'd like to see efforts in both
	halves; I'm interested in both. My number one wish, as far as the
	electroweak sector is involved, is that we
	\textbf{find and are able to study patterns of CP violation in charm
	physics}. As regards the QCD part of things, I really find that
	\textbf{charm has turned out to be a terrific laboratory for QCD
	and its workings}, we're still trying to figure out what's going on:
	all the data's not in yet, and I think we're going to need every bit
	of data to understand what's going on, and that's been a terrific part
	of this workshop. 

	So the completion and growth of those two areas would be what I'd hope
	most for, but by no means saturates what I want to see. Now Bruce has
	handed out to us a bunch of questions that we've gathered from various
	folks; so let me throw it back to Bruce now --- and make sure you have
	a live microphone.
\end{itemize}

\subsection{Mixing: {\boldmath $(x,y)$}, the lattice, and CPV}
\label{disc-mixing}

\begin{chair}%
The first question is a reasonably simple one: 
\question{whether we can firmly establish that mixing occurs --- that
	$x$ and $y$ are not equal to zero --- with (say) $2\,\iab$ of data.
	And going further than that, whether we can determine $(x,y)$
	themselves with twice as much data.}
Anyone want to take that?
\end{chair}

\begin{itemize}
  \item[{[DA]}]
	Sure. If the current central values hold up, then $2\,\iab$ is close
	to sufficient for $5\sigma$ signals in two or three channels. If the
	current central values don't hold up, one might hope that at least
	$4\,\iab$ is \ldots\ so yes, that's certainly the right ballpark.
\end{itemize}

\begin{chair}%
Brian, you don't disagree with that?  Either of the Brians
[Petersen and Meadows], since you're sitting together? No?
\end{chair}

\begin{itemize}
  \item[{[JR]}]
	Just one comment: CDF will contribute to these measurements
	very soon as well. So there will be some help even if the situation
	is marginal with $2\,\iab$: contributions from hadron colliders
	[will help resolve $x$ and $y$].
\end{itemize}

\begin{chair}%
Alright, so the next question may or may not be easy because I
genuinely don't know the answer to it --- I had a pretty fair idea of the
answer of that first one --- and that's the following:
\question{Calculations of New Physics contributions to mixing rely on
	determinations of four-fermion matrix elements: what can we expect
	the lattice to contribute to that?}
So that's a question for Aida.
\end{chair}

\begin{itemize}
  \item[{[AE]}]
	So the methods for calculating 4-fermion operators that contribute
	to \BBbar\ mixing are quite mature and ongoing --- there's a result
	by the HPQCD collaboration using NRQCD heavy quarks --- and in
	principle there's no barrier towards repeating these calculations
	for charm. HPQCD would use the HISQ action, and our collaboration
	[Fermilab] also has an analysis in progress for \BBbar\ mixing,
	and we can certainly do this for the charm sector too. 

	I should add a note of caution however: what we can do with current
	methods are calculations of matrix elements of local operators. So we
	cannot say whether the
	observed \DDbar\ mixing is Standard Model or not, because that
	involves non-local operators --- because you have light particles
	propagating --- and that's something that we can't really calculate
	with current methods. But if one wants to know the expectation values
	of local operators that would come from beyond the Standard Model:
	those can be calculated.
\end{itemize}

\begin{chair}%
OK so let me see if I understand the caveat: What about the extent
to which the observed pattern of mixing comes from the interference of
SM and NP contributions? Does your caveat about long-distance operators
mean that one has a grey area there? Or do they not interfere
because they are short and long distance?\footnote{The discussion moved on 
	from this question, and it wasn't followed up at the time. 
	For the record, the answer is that in addition to the problem
	of hadronic uncertainties in the SM predictions,
	``we do not know the relative \emph{phase} between the SM contribution
	and that from any NP model, so that $x_D$ will lie between the 
	extreme limiting cases of constructive and destructive
	interference.''~\cite{dmix-np}}
\end{chair}

\begin{itemize}
  \item[{[AP]}]
	In New Physics, $\Delta M$ or
	mass-difference comes from heavy intermediate states, and even those
	intermediate states are not present in the SM: squarks, gluinos,
	whatever. So regardless of what happens, at the end of the day
	you get a bunch of matrix elements at the charm scale \ldots\ and
	there are only 8 of them. So regardless of the New Physics model,
	you will get those 8 operators. And so the matrix elements of those
	\ldots

  \item[{[AE]}]
	And they can be calculated \ldots

  \item[{[AP]}]
	Yes.
\end{itemize}

\begin{chair}%
It may not all be this easy. \\[1.0ex]
\noindent
We often say that CP-violation-in-mixing is the
real New Physics signal, and this is part of the progaganda, but up
until now the focus has essentially been on $x$ and $y$.
\question{Do the experimental plans that the existing and future
	collaborations have, reflect the need to measure CPV in mixing?}
Comments?
\end{chair}

\begin{itemize}
  \item[{[DA]}]
	The short answer is of course yes. Every mixing paper in the last
	five years has had CP-violating limits although they've been trivial;
	and one of the most active efforts now of the HFAG-charm mixing group
	is to come up with meaningful CP-violating limits which are actually
	posted on the site, if you choose to go look at it \ldots

	So yes.
\end{itemize}

\begin{chair}%
But that's a nominal yes, isn't it? I understand that (say) for the
$\dz\to\kp\km$ channel we included a CP-violation measurement,
but that was essentially trivial, in that it came almost for free with our
$y_{CP}$ measurement ... aren't there more difficult CPV studies?
\end{chair}

\begin{itemize}
  \item[{[DA]}]
	Well it comes for free with a \dstar\ tag as well, for other
	measurements \ldots
\end{itemize}

\begin{chair}%
Well, at an \epem\ machine it does.
\end{chair}

\begin{itemize}
  \item[{[DA]}]
	Yes. Well, perhaps the fact that the LHCb mixing studies are
	currently CP-conserving might indicate that the experimentalists
	haven't caught up with this \ldots but it's certainly on the agenda.

  \item[{[PS]}]
	Yeah. Our plans have recently changed focus, to look more at CP
	violation. What I was presenting was some of the more complete studies,
	incomplete as they were. And our initial efforts were towards mixing
	measurements. But there's been a fundamental shift in what we've been
	[studying], towards CP violation, and CP-violation-in-mixing.
\end{itemize}

\begin{chair}%
The statement is always in the papers but the emphasis isn't
always in the slides, if you see what I mean, or in people's spoken remarks.
So I guess that's what was behind the question.
\end{chair}

\begin{itemize}
  \item[{[??]}]
	Just related to that, to do the CP violation, at least the direct
	one, you need to get asymmetries also from detector effects and so on,
	so how well are LHCb and SuperB geared towards handling this?
	(And I guess you have similar issues in B-physics, right?)
\end{itemize}

\begin{chair}%
This is actually the next question, or part of the next question, which is
about systematics: 
\question{It's one thing to do a percent-level measurement of mixing,
	but it's quite another to do a $10^{-3}$ or $10^{-4}$ measurement
	of CPV. That's a whole lot more demanding, and it's going to be
	systematics that are the problem.}
Has LHCb thought about this?
\end{chair}

\begin{itemize}
  \item[{[PS]}]
	Yes, but unfortunately we've only begun thinking about systematics.
	We have a lot of ideas on how to evaluate them in the data itself.
	But that's a work in progress.
\end{itemize}

\begin{chair}%
Alan wanted to speak to that.
\end{chair}

\begin{itemize}
  \item[{[AS]}]
	Actually I just had a comment on [the earlier point].
	You have to keep in mind that to see any CP-asymmetry
	effects, you need a nonzero $x$ and $y$. But it's only been since the
	spring, since March, that we've had
	such evidence in hand~\cite{mixing-babar-kpi,mixing-belle-ycp}.
	So that, so
	when you say do experimental plans reflect going after [the CPV phase]:
	since March yes, but before March, even in Belle when we measured those
	things, as you said, they were not the headline measurements,
	because we didn't know what $x$ and $y$ were.

  \item[{[AP]}]
	Well this is true if you stick to CPV-in-mixing. In both of those
	cases. I was actually wondering why the person who
	asked the question wanted to know about time-dependent CP asymmetries,
	I mean of course you have CP-violating asymmetries that are independent
	of $x$ and $y$ --- direct CPV effects ---  and you know,
	those are easy to study, although
	experimental issues are similar; I don't know, if you look at
	\dplus\ decaying to a state that contains charged kaons,
	detector effects for kaons of different charges are significant \ldots

	But the point is that you can have CP violating signals that
	don't have [time-dependence] \ldots

  \item[{[BM]}]
	I just had something ... I think the last mixing paper that I read
	--- experimental paper --- that I read that didn't say anything about
	CP-violation in mixing must have been at least 10 years ago. 
\end{itemize}

\subsection{Rare decay measurements}
\label{disc-rare}

\begin{chair}%
We're going to hop forward to some stuff on rare decays now:
something not unrelated to CPV. This actually cropped up in one of the
talks earlier: that there hasn't been much work at the B-factories on
rare decays. Despite our propaganda, and despite the fact that we put it
in all of our proposals and statements, and all the rest of it --- and I
certainly did, for the charm programme at Belle --- so the question is
\question{what the prospects are for rare charm decays in the rest
	of the B-factory era, or perhaps for a super-flavor factory.}
Why aren't there more rare decay measurements?
\end{chair}

\begin{itemize}
  \item[{[DA]}]
	Everybody complains about manpower, right?
\end{itemize}

\begin{chair}%
[Well I've seen rare decay measurements done,
and then not turned into an ongoing programme, 
because the judgement was that other work would better advance people's careers.
And that judgement is probably correct.]
So, you know, ``manpower'' \ldots\ there are manpower issues
but they're not always just the sheer number of people.
\end{chair}

\begin{itemize}
  \item[{[DA]}]
	To be fair: until rare decays can probe SM sensitivities they
	are just not that interesting. We have a long history of
	\underline{not} finding new physics, and if you want to sign up
	a grad student to work on something where we know he's not going
	to find new physics \ldots\ well we \underline{can} do that,
	[and there are many ways of doing it, but there's little motivation
	for it.]
	But when we do have sensitivity [at the level of the SM predictions],
	I suspect there'll be rather a large line of people who want to
	perform the measurements.
\end{itemize}

\begin{chair}%
But I wonder about this argument, because we've been busy doing
charm physics at B-factories, and I've had guys bowl up to me at
conferences and say
``The beauty sector is the most rich in physics at the moment (which is true),
and offers the best prospects for finding physics beyond the SM at the moment
(which is also true),
and therefore it's \underline{the} most exciting thing,
so why would you work on anything else?
And in particular, why are you working on charm?''
And in the terms in which that question is posed, I don't know that
there is an answer to it. And isn't [the statement about rare decays]
a species of that same argument?
\end{chair}

\begin{itemize}
  \item[{[DA]}]
	Well I'd rather address your first part first:
	Rare decays are much more interesting in charm than they are in B,
	because rare decays actually include CP violation and mixing,
	which are also rare in charm but not in B;
	so from the get-go rare charm processes are a broader field.
	And I'm not actually certain how what I said translated to the
	analogy [with charm-versus-beauty].

  \item[{[KS]}]
	As long as you have this up, I have a question for David Asner.
	When you were comparing what could be done at LHCb
	and the super-B-factories,
	most of the time the things were sort-of comparable,
	but there were a couple of instances in which you said you really
	need SuperB for that \ldots\ Now I have to ask you a question whose
	flavour is completely different from [that of] your talk.
	Does one, for those rare cases for which SuperB is essential,
	spend \$500 million?

  \item[{[DA]}]
	That's a good question: it depends what part of the programme of
	SuperB you think is most important.
	If you are a tau physics enthusiast, there is no comparison between
	the tau physics programme at SuperB and the tau physics programme
	at LHCb. [And for] anything that requires a primary vertex
	(including, probably, charm semileptonic decays) the performance
	will simply be better at SuperB.
	To be fair, I've probably represented LHCb in a rather favourable light.
	After all, it's going to be an existing experiment.

	Will the world end if we don't have the charm physics programme from
	SuperB? No. We're not trying to sell SuperB on its charm physics
	programme. But if you can motivate SuperB based on the way that
	the B-physics interacts with the LHC programme, then you get
	the charm and the tau and the charmonium and the ISR programmes
	\emph{etc.}\ for free, assuming that you can assemble the
	workforce.

  \item[{[KS]}]
	But if there's one thing I have learned, it's that when someone is
	blessed with huge luminosities, they have the advantage of making cuts
	and throwing away so much data that they can [do almost anything].
	I always used to say that the B-factories, if they wanted to get at
	whatever I might be doing, they can cut everything out and still beat
	us on total statistical and other kinds of errors.

	And that is what makes me wonder about comparison between LHCb and the
	SuperB. LHCb will have so much production of whatever-you-want,
	that you wonder whether they have the ability to throw away a lot of it,
	and become, you know \ldots

  \item[{[DA]}]
	Well they do throw away a lot of events: it's called the trigger, right?

  \item[{[KS]}]
	Yes, the trigger, and whatever else you want \ldots\ and still reach
	the statistical precision [of SuperB].
\end{itemize}

\begin{chair}%
There are also issues of systematics, remember: things that a hadron
experiment by its nature doesn't do as well. Alan?
\end{chair}

\begin{itemize}
  \item[{[AS]}]
	There are many final states that would be very challenging at a
	hadron collider: $\pi^0$ and $\eta$ and charged $\rho$, you know,
	those will be very difficult at a hadron machine.
	You will not measure $\bz\to \pi^0\pi^0$ at LHCb --- I don't think ---
	and I don't think you'll see $\bz\to \rho^+ \rho^-$;
	and many of the $\eta$ and $\eta'$ decays will be very hard.
	I mean, obviously they can be simulated, but there are many many
	final states where I don't know how it's going to happen
	in a hadron environment.

  \item[{[BM]}]
	Just one thing to add to what Alan was saying: One thing you
	probably want to do in looking for CP-violation in mixing is, you want
	to see if there's a difference in $x$ and $y$ for \dz\ and \dzbar.
	And then you want to see if they're the same in different decay modes,
	in all the different decay modes.
	And I think that at SuperB that will be much easier to do.
\end{itemize}

\subsection{Do we claim too many new states?}
\label{disc-newstates}

\begin{chair}%
I'm going to move on, to something that's going to induce a
different kind of argument, but this is also about perceptions of what
is exciting and what's not. One of our questioners wanted to know
\question{whether we are claiming too many new states,
	either quarkonium-like or in general.
	And what about dynamics? Particularly near threshold,
	since a lot of these states do seem to be near threshold, \ldots}  
An awful lot of new states are being claimed,
some of them on our experiments' websites.
[I think Yifang has his hand up to answer this.]
\end{chair}

\begin{itemize}
  \item[{[YW]}]
	I think, in fact, we do see a lot of resonances, or particles, or
	structures --- whatever you want to call them --- near the threshold,
	and I think there's a very simple experimental reason,
	in that near the threshold the background is much lower,
	so wide resonances are much easier to see \ldots
\end{itemize}

\begin{chair}%
OK, but let me pick you up on the words that you used there: 
yes it's easy to see ``it'', and you said ``resonances'', ``new particles'',
or ``structures'', but just because you see nontrivial structure,
it doesn't mean that there's a new pole sitting [on the complex plane].
\end{chair}

\begin{itemize}
  \item[{[YW]}]
	I agree: I said all these possibilities, because now,
	particularly at BES [where there are so many things that we
	see], we have a lot of arguments about whether new particles or
	structures or enhancements (or whatever) exist,
	and I think shows that we need more statistics,
	and we need a much better detector.
\end{itemize}

\begin{chair}%
No, but hang on: don't we also need better interpretation?
\end{chair}

\begin{itemize}
  \item[{[YW]}]
	Of course, later on; I think the important thing, right now, if we
	see a signature of (say) $5\sigma$, $4\sigma$ sometimes, $6\sigma$,
	then many people believe it, some people don't believe it, 
	and for both experimentalists and theorists, [the confusion] 
	makes it hard to work together. 
	Now if you have a much better detector,
	and significantly improved statistics,
	if you see a signal at a level of say $20\sigma$, or $50\sigma$, then
	there's no question of whether these things exist or not, and it makes
	it easier for both experimentalists and theorists.
\end{itemize}

\begin{chair}%
Perhaps it will clarify the point [if we consider another example. 
There have been many claims of new states at Belle,
and one of the most interesting pieces of recent work concerns 
the extra enhancement below the 4260, in the $\epem \to \pi^+ \pi^- \psi$
cross-section. Yes? I think everyone has seen this.
Now in the paper~\cite{y4260-belle}
we were very careful not to say ``this is a new state''.
But in \texttt{arXiv} postings and on the web, it is being given a name,
a ``Y'' and a number, with it being implied or stated that it's a new state.]
And that reflects the fact that for (I think) 80\%
of the people in the community saying that you see a significant
enhancement --- and it \underline{is} an enhancement,
		and it \underline{is} significant, OK,
		that part's not in question: 
		a better detector won't change that ---
but the first thing we always lean to is, ``it's a new state''. And that's
been the first interpretation rather than the last one, for lots of
things from Belle, from BaBar, from BES, and what this question is asking is,
\question{Is that really a good thing --- and if it's not,
	how do we get past it?}
And Estia wants to speak to that.
\end{chair}

\begin{itemize}
  \item[{[EE]}]
	I mean, when you have wide structures, it's always rather difficult
	to figure out what's going on, and you can certainly show examples
	where a single pole will give rise to complicated-looking structure. 

	So there's two separate questions. For the theorists, they have to
	figure out how to interpret the structures that are seen in experiment,
	in terms of what's underlying them. I don't know how the experiments
	can do that, but what experiment can certainly do is describe
	the properties of those structures --- make sure you understand
	its quantum numbers, if you interpret it as a state,
	its decay properties \emph{etc.}:
	these things will certainly make it a lot
	easier for theorists to try to disentangle wide structures into what
	actually is underlying them [and whether] there are additional states.

	Every bump is not a new state. You can show examples when it is
	clearly not \ldots
\end{itemize}

\begin{chair}%
And was it you that showed that point about the $\pi^+ \pi^- \psi$ and the
$\pi^+ \pi^- \psi'$ [cross sections, in ISR data from the B-factories],
and that maybe \ldots
\end{chair}

\begin{itemize}
  \item[{[EE]}]
	Yes, it may very well be that there's one state with different
	decay modes at different energies, whose branching fraction varies as a
	function of energy. Now, I can't calculate that one, because I don't
	know what it is --- if it's a hybrid of some sort, then I don't know
	how to do the calculation of how much it'll decay into those two modes
	--- but that's (in a sense)
	a theorist's or a phenomenologist's job to try to figure that all out,
	but experimentalists can certainly try to disentangle any other decay
	modes that are in the same region, for example,
	besides the $\psi$ or the $\psi'$,
	maybe there's another mode in between that can fill us in on what's
	going on. But that's all an experimentalist should [be expected to do].
	The interpretation is difficult.

  \item[{[EG]}]
	This goes back as long as I can remember in hadron physics:
	the $\sigma(600)$ is something that people call different things
	\ldots\ I don't know what to call it, but if I do know how to use it,
	you can call it whatever you want.

  \item[{[AP]}]
	There's actually a little counter-example to that: You all know the
	story of the pentaquark, which was very well predicted by several
	groups, and very well observed by several different experiments, and
	yet's it's gone. And so it's natural that you see so many states, well
	not states, but structures. Time passes, and you sort out which of
	those are real states,
	and which of those are just threshold things \ldots

  \item[{[YW]}]
	I think I haven't finished what I want to say, so let me finish, and
	then we can go forward. I think it's very important to have more data.
	If you have $4\sigma$, $5\sigma$, how do you disentangle whether
	it's a real particle, structure, a fake, or so on?  But once you
	have significantly improved statistics, you can certainly establish
	whether this is a real structure, and you have the possibility to
	understand its decay properties, dynamics and so on, and then
	you can have a much better understanding.

	Put it this way: up to now, we have studied a lot of electroweak
	physics, but we understand QCD much less well. And we know that
	we should see a lot more resonant states, hybrids or whatever,
	glueballs, beyond the conventional \qqbar\ or baryon states:
	certainly more than what we see.

	So with significantly improved statistics \ldots\ put it this way,
	we now see many $4\sigma$ signals. Over time, BES-III will increase
	statistics by a factor of a hundred, and on top of that the detector
	is much better --- corresponding to an improvement, for the same
	statistics, by a factor of two or four --- so this is an improvement
	of almost a thousand in total.
	So you can imagine: with one thousand more better signals, I think
	it will be revolutionary, and for many unresolved problems or questions,	[it will give us] a much better understanding.
	So I think it's important to have more data.
	And by that time, once we have firmly established particles
	and their propeties, theorists can work together, and figure out all
	these details. And [regarding] QCD: by that time, we'll have a much
	better understanding.
\end{itemize}

\subsection{Comparing the lattice to experiment}
\label{disc-lattice-comparison}

\begin{chair}%
We're going to move on. This is a question I think you'll even like:
\question{Suppose experimental and lattice values for decay constants
	disagree by $2\sigma$.
	Is that a discrepancy, and should we be worried about it?
	And if $2\sigma$ is not, then what about $3\sigma$ or $4\sigma$?} 
[To AE:] Do you have a a view from your side?
\end{chair}

\begin{itemize}
  \item[{[AE]}]
	I would say that at present, the situation is not yet completely
	clear, whether or not there is indeed a discrepancy.\footnote{The
		reference is to the disagreement in $f_{\ds}$ 
		between
		the new lattice calculation $(241\pm3)$ MeV~\cite{fds-hpqcd}
		and the CLEO-c average $(275\pm 10 \pm 5)$ MeV
		presented in Steven Blusk's talk at the workshop.}
	I'd also like to
	point out that the experimental values for the decay constants have to
	assume a value for $V_{cs}$, which is not very well determined
	at this point. It's hard to imagine right now that $V_{cs}$ could
	change too much, but you can have at present a few small effects which
	would bring experiment and theory back [together]. I think once the
	error bars decrease more, then we would be more worried.

	To really do a precision test of lattice QCD without any assumptions
	about CKM angles, we need to compare the CKM-free quantities
	like the ratio of the semileptonic decay rate
	to the leptonic decay rate, where the CKM angle does not contribute.
	So [at the level of precision reached by HPQCD]
	I don't think that just looking at the decay constant itself is
	really [enough]: you want to look at the ratio of semileptonic decays
	to leptonic decays;
	you want to look at the shapes of the form factors;
	you want to see if you extract $V_{cs}$ from the semileptonic,
	does that agree with the result from the leptonic case,
	to see if there really is a discrepancy. So I think we need
	more information than just the decay constant.
	And I think that's part of the programme.
\end{itemize}

\begin{chair}%
And that's a point where Yifang's argument about overwhelming
statistics is important, because with overwhelming statistics you have
the luxury of doing those assumption-free tests that you're talking about.
\end{chair}

\begin{itemize}
  \item[{[AE]}]
	That's certainly true. I mean, two sigma statistical fluctuations
	are not unheard of;
	and while Peter Lepage might vehemently disagree with me for
	saying this, it's not absolutely impossible that (say) the HPQCD
	result might have slightly larger errors than they [claim],
	in which case the discrepancy is not as large as it currently is.
	So I think it's certainly important to monitor and keep in mind,
	but ah \ldots
\end{itemize}

\begin{chair}%
Does anyone else want to speak to that? It's a fairly complete
answer, but \ldots
\end{chair}

\begin{itemize}
  \item[{[AP]}]
	I'd like to ask a question about the errors that lattice
	calculations are assigning to their numbers. I mean, I used to see nice
	lattice predictions with very small error bars that were completely
	overtaken by new lattice predictions, and people would say that they're
	inconsistent, but ``That's OK, because the previous error bar did not
	include quenching errors.''
	Now, how can one claim an error bar if there is an error bar
	sitting on top of that, which you don't even know what it is?
	Now I understand that with the new techniques this [is no longer
	the case]. But when you now do things at the level of a percent,
	then none of these errors, they're not independent,
	and you have to take into account correlations between them.

  \item[{[AE]}]
	And we do. And we do. We have learned a lot in the last ten years on
	this. I want to make this point very clearly, and I think it requires
	non-lattice theorists and experimentalists maybe to have some
	knowledge, but if you read a paper and it says we are doing
	a quenched calculation and we get this result with this error,
	just, don't even \ldots\ they did an error analysis \ldots
\end{itemize}

\begin{chair}%
But it's quenched.
\end{chair}

\begin{itemize}
  \item[{[AE]}]
	But it's quenched. And even if they have some estimate for the
	quenching error, but if it's not anywhere close to 20\% or so,
	you just shouldn't believe it.
	The same thing with calculations that have two flavours of sea quarks,
	that's not enough: there are three in the real world.
	So, you know, I don't think it makes a whole lot of sense to
	compare quenched lattice calculations to experiment.
	I also don't think it makes a lot of sense to compare
	lattice calculations based on two flavours of light sea quarks
	to experiment. [These comparisons are not interesting to people 
	outside of lattice field theory.] And that's the reason I only showed
	results based on $2+1$ sea quarks.

	And the second point to your question, it's a valid point:
	we have a long history as a community in lattice field theory,
	certainly [more than] 10 years ago people were not worried about systematic errors
	and as we got more sophisticated we became more worried.
	And not everyone in the field understands that they need
	to be very careful with the systematic error analysis.
	But I think that people who \underline{are} claiming
	to do serious calculations need to be able to show you plots:
	Do you understand the light quark mass dependence of your result?
	Have you looked at the lattice spacing dependence?
	I mean you need to have a very sophisticated systematic error analysis,
	and if you don't, you shouldn't believe the result.
	[Certainly not at the few percent level.]

	And it's always good to have more than one person,
	because these analyses are complicated. I mean more than one group
	--- one person never does these analyses by themselves ---
	but more than one group do these calculations with different
	methodologies and see that you have that consistency.
	So, you know, we have the HPQCD result, we have our result,
	which is a factor of two less precise than their result,
	but within \underline{our} error, the two results are
	certainly very consistent.

  \item[{[DA]}]
	Can I jump in real quick?
\end{itemize}

\begin{chair}%
Quickly \ldots
\end{chair}

\begin{itemize}
  \item[{[DA]}]
	Speaking to the experimental determination of the CKM-independent
	quantities, at best I can tell from the two reports I mentioned
	earlier, it looks like the measurement is at about the 10\% level.
	Part of the BES-III programme will get that to, you know, a few percent,
	and SuperB will be required to get 7\%, but these are part of \ldots
\end{itemize}

\begin{chair}%
We might come back to that later.
\end{chair}

\subsection{BaBar data after datataking ends}
\label{disc-babar-comparison}

\begin{chair}%
Before Gene goes, I just thought
we'd ask the question you see here:
\question{whether there will be an analysis effort on BaBar beyond 2009.
	And what the plan is for doing such an obviously difficult thing?}
I mean, we have some BaBarians or ex-BaBarians here in the room:
anyone want to speak to it?
\end{chair}

\begin{itemize}
  \item[{[??]}]
	As far as I understand, there will be three years of \ldots\ at least
	the infrastructure for doing analysis for three years after datataking.
\end{itemize}

\begin{chair}%
What does ``infrastructure'' mean?
\end{chair}

\begin{itemize}
  \item[{[??]}]
	All the computing, so you can process all the data, do the
	simulations, and put new decay modes in.
\end{itemize}

\begin{chair}%
Right, I mean, if you want to do ten decay modes, will you have a
supply of slaves to do it, or will you have to do it all yourself?
\end{chair}

\begin{itemize}
  \item[{[??]}]
	Clearly, there is a very limited number of new graduate students
	coming on, already now.
	So, \ldots\ they will be graduating in the three years \ldots
\end{itemize}

\begin{chair}%
So what's being done to prioritise measurements?
\end{chair}

\begin{itemize}
  \item[{[??]}]
	We have a core list of results which were scheduled to come out \ldots\ 
	somewhere next year or the year after that. And they are being
	prioritised within B-physics and charm physics, so for instance, the
	mixing analysis will be updated with the full dataset. So there is a
	list of people, with their names assigned to core analyses \ldots
\end{itemize}

\begin{chair}%
I don't know, does someone who's lived through this part of the
life-cycle of a big experiment want to speak to this question?
\end{chair}

\begin{itemize}
  \item[{[DB]}]
	In addition to that, there is also serious consideration being given
	to conserving the data for future use well beyond 2011, because it is
	recognised \ldots
\end{itemize}

\begin{chair}%
This goes to the second question posted here actually,
\question{what to do when the data becomes a heritage matter.}
\end{chair}

\begin{itemize}
  \item[{[DB]}]
	I'm not aware that there's a solution to that yet, but certainly
	that question is being considered.
\end{itemize}

\begin{chair}%
My question is: when it's several years out from the running of the
experiment and you've got one or two dedicated groups at a slow burn
working on this data, how [do] you maintain quality? How do you stop
people with agendas and nothing better to do, and a lot of BaBar data \ldots
\end{chair}

\begin{itemize}
  \item[{[KS]}]
	You don't avoid that.
\end{itemize}

\begin{chair}%
That just happens?
\end{chair}

\begin{itemize}
  \item[{[KS]}]
	It's natural. Things die, and there's a rate of dying, and then
	there is that not that close supervision, and the quality suffers.

  \item[{[AP]}]
	This might open a can of worms, but are there any plans to make the
	data publicly available on the internet, just like astronomy?
\end{itemize}

\begin{chair}%
But that then multiplies the issue that I was just raising, doesn't
it. I mean, if some high-school student can run a job can you imagine
the number of new states we're going to be seeing \ldots [laughter]
\end{chair}

\begin{itemize}
  \item[{[AP]}]
	But what's wrong with that? It's not going to be signed
	by BaBar \ldots 

  \item[{[DB]}]
	\ldots\ but the answer is yes.
\end{itemize}

\begin{chair}%
[Gene leaves]
\end{chair}

\subsection{Fermilab: charm, and antiprotons}
\label{disc-fermilab}

\begin{chair}%
We've perforce been skipping around here, so I'll just continue on
this slide for a moment, 
\question{[can we hear] what the plans are at Fermilab for charm studies
	--- at D0 and CDF?}
\end{chair}

\begin{itemize}
  \item[{[JR]}]
	[inaudible] certainly mixing is going to be out soon, with
	comparable sensitivity to \ldots
\end{itemize}

\begin{chair}%
OK, I already knew the answer to that question, but what about
non-mixing topics?
\end{chair}

\begin{itemize}
  \item[{[JR]}]
	I have no good overview, except for charm production \ldots 
	[from elsewhere: ``and the rare decays''].
\end{itemize}

\begin{chair}%
\ldots\ and the rare decays. I guess the question is whether
there will be an update or successor to that analysis,
or whether that's the last thing we'll see.
[to AP: ``You have a lot of comments!'']
\end{chair}

\begin{itemize}
  \item[{[AP]}]
	No, I just happen to be at an institution that participates in CDF,
	and which actually does rare-decay charm analysis.
	So, you know, the first paper from CDF-II
	was on $D^0 \to \mu^+\mu^-$~\cite{mumu-cdf}
	\ldots\ and right now they're doing an update of that.
\end{itemize}

\begin{chair}%
There was a second question posted here --- it's outside my
awareness --- about
\question{what the support in the community is for the antiproton source
	proposal at Fermilab.}
Does anyone want to comment on that?
\end{chair}

\begin{itemize}
  \item[{[DB]}]
	I can say my personal opinion --- I cannot speak for the community,
	of course --- because I have been following a little the proposal for
	the antiproton source, to make a dedicated experiment at the pbar
	accumulator --- my personal impression is that the next step, the next
	generation of hadronic \ppbar\ physics will have to be a global
	approach like PANDA, with a detector like PANDA.
	A detector like we had in E835 has more-or-less said whatever
	it could say, and I don't know whether Kam would agree,
	but we were discussing the other day: already
	at the end we were sort of at the limit of what could be done in this
	kind of experiment.

	So I don't know what was the point of the question:
	my impression is that there is very little one can do in the limited
	space which is available there, unless one can dig a bigger hole,
	and really do a general-purpose detector.

  \item[{[KS]}]
	As a former worker in the field: I can tell you that there
	is just not enough constituency and it doesn't rank high enough in the
	priority, so at least my feeling is that it is very sad that the only
	place which has currently the ability to produce antiprotons is not
	planning [anything for the future].
	But that's the way life goes: GSI is coming,
	[the current facility has] outlived its capabilities, and I
	don't see much future for that at Fermilab.
	That is my personal opinion,
	but I make that opinion after having heard the arguments
	and the things that have been proposed.
\end{itemize}

\begin{chair}%
OK, I'll take that as an answer. I'm going to skip to the second part ...
\end{chair}

\begin{itemize}
  \item[{[DA]}]
	About SuperB?
\end{itemize}

\begin{chair}%
Yep. [i.e.\ \question{What is the interest and the support of our community
	for the Super-B project?}]
\end{chair}

\begin{itemize}
  \item[{[DA]}]
	OK I'll start with a question to Alan --- or maybe to you, Bruce ---
	as to how many names are on the Belle letter of intent for SuperBelle?
\end{itemize}

\begin{chair}%
At least half [of the collaboration].
\end{chair}

\begin{itemize}
  \item[{[DA]}]
	Because SuperB has three-hundred-and-twenty signatories, and there's
	some overlap between the two, but it's not overwhelming [\emph{i.e.}\ 
	the overlap is not the dominant part], but we're talking about hundreds,
	five or six hundred people in our field, theorists and experimentalists,
	who are excited to do work, who have produced these documents \ldots\ 
	so I think there is considerable support for \underline{a}
	super-B-factory.

  \item[{[YW]}]
	I'd like to say that although we are not part of the super-B
	[project, we strongly support it] \ldots\ we think it's very good
	for the field, and very good for the community.
	Without super-B \ldots\ at least for us, we at BES,
	five years from now, the charm field would only be at BES,
	and um, we don't want that.
\end{itemize}

\subsection{What to measure at LHCb}
\label{disc-lhcb-measurements}

\begin{chair}%
There was a pessimist's question on exactly this point, which you
saw, which I'm not going to put up, because we're running out of time.
[Instead I want to consider] a couple of LHC-related questions ---
essentially LHCb --- the first one was answered by Patrick's talk; so
the second one is
\question{whether we should be making a shopping-list of the
	measurements that we want LHCb to make}
--- the people in this room, this community --- how much [would that help?]
\end{chair}

\begin{itemize}
  \item[{[PS]}]
	I think that that's an excellent idea. People complain about the
	manpower issue at the B-factories: at LHCb it's even worse right now.
	So having a list of things to look at --- and a prioritised list --- 
	would be helpful for our planning.
	As scientists come off of the building phase of the experiment
	they're going to be looking for analyses. And if we have an enumerated
	list of charm physics [topics] we might be able
	to woo them into this field. 
\end{itemize}

\begin{chair}%
OK, so a specific question, as to
\question{whether there's a charmed baryon programme at LHCb.}
\end{chair}

\begin{itemize}
  \item[{[PS]}]
	Right: Not now. Not yet.
\end{itemize}

\begin{chair}%
Will there be?
\end{chair}

\begin{itemize}
  \item[{[PS]}]
	I don't know. It would take more than just me to be doing it.
\end{itemize}

\begin{chair}%
OK, but realistically: would the people in this room saying ``We want
such-and-such to be measured'' make a difference to that, or is it just
the people in the collaboration?
\end{chair}

\begin{itemize}
  \item[{[PS]}]
	I think it's more getting individual scientists and groups in the
	collaboration interested in that work. 

  \item[{[JR]}]
	I just wanted to second that. I think that at the moment, suggesting
	exciting analyses might actually still have an influence, for the very
	reason that many people who have worked very hard on building the
	detector are moving into physics at this moment, and we can have
	influence over [the choices that they make].
\end{itemize}

\begin{chair}%
Maybe we should look into the list idea.
\end{chair}

\subsection{On being a successful sideline:}
\label{disc-sideline}

\subsubsection{ISR work at a ``super-B-factory''?}
\label{disc-sideline-isr}

\begin{chair}%
Going to another sideline point about the future, this is one for
SuperB: the ISR programme at the B-factories has become quite an
important part of the work lately, and through accumulated accidents
that kind of dropped out of this meeting.
[But there's good physics being done using ISR at present.]
So, 
\question{what ISR programme is foreseen for the super-B 
	or super-flavor factories?}
\end{chair}

\begin{itemize}
  \item[{[DA]}]
	There are a couple of pages in the SuperB CDR precisely on this
	physics, and I can tell you that at every SuperB meeting that is about
	the physics (not the accelerator) there are talks on this subject. So
	it's on the radar. The data will get read out: it will be there on tape.
	But it will require people who are interested in this physics from the
	current experiments showing up and doing that physics \ldots\ I don't
	think the ability to do this physics, as interesting as it is,
	will drive whether or not a super-B machine gets built \ldots
\end{itemize}

\begin{chair}%
No, I think \underline{that}'s right \ldots\ Because it took us a long time to
warm up to doing ISR physics. It was in the BaBar book --- we didn't
have a book, but it was in the BaBar book --- but it even took them a
while to warm up to it, so I guess that's what's behind the question.
\end{chair}

\subsubsection{Charm studies at LHCb?}
\label{disc-sideline-charm}

\begin{chair}%
Some general questions about the future now: This is actually a question
that I care about as well, as to
\question{how easy it's going to be doing charm
studies parasitically at facilities that aren't dedicated to it.}
This is really an LHCb question: we managed it fine at Belle and BaBar, but
Belle and BaBar run with open triggers. Now ...
\end{chair}

\begin{itemize}
  \item[{[PS]}]
	Yes. We are in some sense competing for bandwidth. And we may have
	to make a case for every exclusive charm channel that we put in, and
	certainly it needs to be optimised so that it doesn't eat into
	the total bandwidth too much.

	But I think that anything that someone is genuinely interested in doing
	can be put into the trigger at this stage. It's more of a manpower
	issue, even at this stage. And if I may make a comment about the poor
	neglected LHCb upgrade --- I'm sorry I didn't mention anything about
	that --- the trigger is going to have to be completely overhauled
	for an LHCb upgrade in order to accomodate the increased luminosity,
	and although it's by no means finalised,
	the prevailing idea for the trigger there
	is to go with a full software trigger. And once you've done that,
	it opens up the whole field: if you can efficiently create a charm
	trigger, you don't have these $p_T$ cut limitations that are applied at
	the hardware level. 

	So, yes. It's a matter of manpower and interest and influence, I think.
\end{itemize}

\begin{chair}%
Anyone else on that one? No?
\end{chair}

\subsection{Charm \& the heavy ion community}
\label{disc-heavy-ion}

\begin{chair}%
OK this is a question that I wasn't
expecting, but it's a potentially useful one, and it's
\question{whether a closer collaboration with the heavy ion community
	would be beneficial, either for charm or quarkonium physics.} 
Anyone with experience of heavy ion work? Or friends in heavy ion work?
\end{chair}

\begin{itemize}
  \item[{[AP]}]
	There is already close collaboration. First of all, they are already
	going to share an accelerator: the LHC. And you know, those guys need
	fragmentation functions, things like that: they get it from us. 

  \item[{[KS]}]
	Already at Brookhaven, people are beginning to do spectroscopy,
	which was not their [original] goal.  So they are of necessity coming
	into the field, and they often ask questions: it is not quite a
	collaboration yet, but yes, each is beginning to get interested
	in the other. I think it's a developing thing.
\end{itemize}

\begin{chair}%
Right, I assume that's the thought behind the question: that the
communities are somewhat disjoint.
\end{chair}

\begin{itemize}
  \item[{[KS]}]
	At RHIC it's beginning.
	At LHC it remains to be seen, because it is still in the future.

  \item[{[AP]}]
	There are already a couple of groups: I mean, at our university,
	Michigan State, the groups are both \ldots\ 
	there are groups on both CDF and D0, CLEO, and STAR and PHENIX
	and so people are \ldots\ maybe it needs to be formalised.

  \item[{[KP]}]
	Also at GSI with the FAIR project we have two distinct communities
	\ldots\ [but] it's exactly ten metres from my office to [that of]
	my colleague who leads the heavy ion community there, 
	and our intention is to place an application for a virtual institute 
	\ldots\ to manifest [our cooperation] a little more: to really
	concentrate on charm and the various issues with the nucleus
	\ldots\ things are moving together to some extent.
\end{itemize}

\subsection{On charmed baryon measurements}
\label{disc-baryons}

\begin{chair}%
OK. We are \ldots\ running out of time. I thought I'd just throw one
slightly odd question in. There was a comment along the way
\question{about the measurements that are being made in baryons,
	and that there appear to be some kind of imbalance in it:}
that you can have some of these quite
sophisticated studies that measure the spin and parity of some of the
baryons, or search for new states or whatever, but there are quite
fundamental things that we know poorly, particularly about the $\Omega_c$.
And some of those measurements aren't being made. Is there a way around
that? Or do we need a super-motivated group to just do the work at the
B-factories? I mean, CLEO-c was going to run at $\Lambda_c$ threshold at
one point, and famously \ldots
\end{chair}

\begin{itemize}
  \item[{[DA]}]
	SuperB --- the INFN proposal --- has plans to spend a month at the
	$\Lambda_c$ threshold. But it's \underline{at} $\Lambda_c$ threshold. 
\end{itemize}

\begin{chair}%
Right: they talk about it, but CLEO-c did as well. Is it going to be
the first thing that gets dropped, in the same way that \ldots
\end{chair}

\begin{itemize}
  \item[{[DA]}]
	I think the answer is, ``of course''.
\end{itemize}

\begin{chair}%
Right. OK. That's not necessarily a criticism \ldots
\end{chair}

\begin{itemize}
  \item[{[DA]}]
	The entire 4 GeV programme would probably be the first thing to get
	dropped if the 10 GeV luminosity wasn't up to snuff.

  \item[{[KP]}]
	To give you one example, our group started the idea to measure \ldots\ 
	the $D_{s1}(2536)$ which we have seen this morning~\cite{ds1-babar};
	and it really has reached excellent resolution \ldots\  
	But to give you an idea of the timescales, this PhD student has spent
	three years on it, to get really all the systematics nailed down
	to that point where you can reach these 100 keV systematic errors 
	\ldots\ and that's exactly what kills a lot of these mass measurements.
	If there is not really an important issue,
	you just drop it because these systematic studies \ldots\ 
	there are so many other things you could do, which are more
	geared to the mainstream. 
\end{itemize}

\begin{chair}%
Less mainstream and more difficult.
\end{chair}

\begin{itemize}
  \item[{[KP]}]
	Yes.
\end{itemize}

\subsection{Thanks}
\label{disc-closing}

\begin{chair}%
OK look, I ... don't want to keep you here any longer, and I think
we've covered a fair amount of territory on a range of things. So thanks
to the people who served on the panel, and to all of the speakers I
guess. And um, I would certainly like to thank the committee, for what's
been a good workshop.
\end{chair}

\bigskip 

\end{document}